УДК (930.1/477):(2.67/524.8)

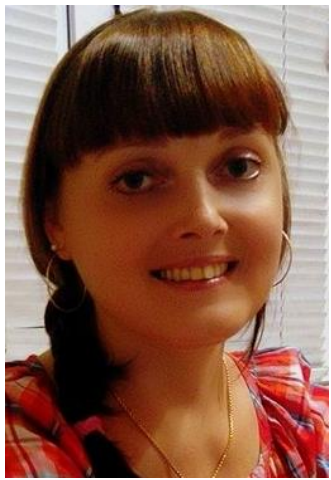

**КОЛТАЧИХІНА
Оксана Юріївна,**
кандидат історичних наук,
Кфар Сільвер вища школа – академія
Наале, викладач фізики,
oksana.koltachykhina@gmail.com
Ізраїль, Україна

# ВПЛИВ ПРАВОСЛАВНОГО ХРИСТИЯНСТВА НА КОСМОЛОГІЧНІ ІДЕЇ В УКРАЇНІ (XI–XVIII ст.ст.)

*Вперше детально описано вплив православного християнства на розвиток космологічних ідей в Україні протягом XI–XVIII століть. Досліджено еволюцію існуючих у той час моделей про народження та будову Всесвіту. Показано зв'язок православних ідей з космологічними уявленнями в Україні. Проаналізовано маловідомі праці перших викладачів українських академій з досліджуваної тематики. Встановлено, що на теренах України були відомі всі моделі Всесвіту, проте тексти православних богословів мали великий авторитет і вплив на розвиток космологічних ідей.*

*Проведене дослідження дало можливість визначити специфіку вітчизняної культури в загальноосвітній культурі. Розглянута еволюція ідей, показала й історію мислення та розвиток свідомості українського суспільства. Крім того, публікація несе й пізнавальний аспект.*

*Статтю виконано у рамках проекту «Наука та православ'я в світі» Інституту історичних досліджень Національного фонду досліджень Греції, що стало можливим завдяки підтримці гранту Всесвітнього благодійного фонду «Темплтон». Висновки автора статті можуть не відображати точку зору Проекту та Всесвітнього благодійного фонду «Темплтон».*

***Ключові слова:*** *православ'я, космологія, Україна, Всесвіт, космологічні моделі.*

# THE INFLUENCE OF ORTHODOX CHRISTIANITY
# ON COSMOLOGICAL IDEAS IN UKRAINE (XI–XVIII centuries)

*For the first time in detail the influence of Orthodox Christianity on the development of cosmological ideas in Ukraine during the XI–XVIII centuries is described. The evolution of existing models of the birth and structure of the universe at that time was investigated. The connection between Orthodox ideas and cosmological notions in Ukraine is shown. The little-known works of the first teachers of Ukrainian academies on the studied topics are analyzed. It was established that all the models of the universe were known in Ukraine, but the texts of Orthodox theologians had great authority and influence on the development of cosmological ideas.*

*The conducted research made it possible to determine the specifics of domestic culture in the general education culture. The evolution of ideas has been considered, and has shown the history of thinking and development of consciousness of Ukrainian society. In addition, the publication also has a cognitive aspect.*

# ВЛИЯНИЕ ПРАВОСЛАВНОГО ХРИСТИАНСТВА
# НА КОСМОЛОГИЧЕСКИЕ ИДЕИ В УКРАИНЕ (XI–XVIII вв.)

*Впервые описано влияние православного христианства на развитие космологических идей в Украине в XI–XVIII вв. Исследована эволюция существующих моделей рождения и структуры Вселенной в то время. Показана связь между православными идеями и космологическими представлениями в Украине. Проанализированы малоизвестные работы первых преподавателей украинских академий по исследуемой теме. Установлено, что все модели Вселенной были известны в Украине, но тексты православных богословов имели большой авторитет и влияние на развитие космологических идей.*

*Проведенное исследование позволило определить специфику отечественной культуры в общеобразовательной культуре. Была рассмотрена эволюция идей и показана история мышления и развития сознания украинского общества. Кроме того, публикация также имеет познавательный аспект.*

*Статья была реализована в рамках проекта «Наука и православие во всем мире» Института исторических исследований Национального фонда эллинистических исследований, что стало возможным благодаря поддержке гранта Всемирного благотворительного фонда «Темплтон». Мнения, выраженные в этой публикации, принадлежат автору и не обязательно отражают взгляды Проекта и Всемирного благотворительного фонда «Темплтон».*

**Ключевые слова:** *православие, космология, Украина, Вселенная, космологические идеи.*

Пам'ятники стародавньої української думки являють собою невід'ємну частину органічної історії та культури нашого народу. Їх дослідження дає можливість визначити специфіку вітчизняної культури в загальносвітовій культурі. Історія культури це не лише історія ідей, але й історія самого мислення та еволюції свідомості людини. Особливу увагу дослідників, істориків науки потребують питання зв'язку релігії та науки адже вони є формами духовно-творчого освоєння людиною світу. Співвідношення релігії та науки є актуальним у даний час. У попередніх дослідженнях автор статті показала історію розвитку уявлень про світ, що були розповсюдженні на теренах сучасної України [1, 2]. У представленій роботі зроблено спробу показати зв'язок православ'я з космологічними уявленнями в указаний період.

Намагання розв'язати поставлену проблеми здійснено в низці праць з історії розвитку релігії, освітніх закладів та окремих наук, зокрема фізики й астрономії [3–11]. Так, у «Нарисах розвитку вітчизняної астрономії» подано уявлення про будову Всесвіту стародавніх народів, зокрема тих, що населяли землі сучасної України [5]. У монографії «Природознавство в Україні до початку XX ст.» розглянуто історію природознавства в контексті соціокультурного процесу в Україні, на тлі загальносвітового цивілізаційного поступу [6]. Монографія складається з шести розділів, де подано історію розвитку наукових знань у ту чи іншу історичну добу.

Однією з перших спроб встановлення питання сприйняття геліоцентричної системи в Росії була праця Б. Є. Райкова «Нариси з історії геліоцентричного світогляду в Росії. З минулого російського природознавства» [7], де подано також фрагментарні відомості про Україну. Цікавим історіографічним джерелом є праця М. В. Головка «Використання матеріалів з історії вітчизняної науки при вивченні фізики та астрономії» [8]. Автор коротко висвітлив наукові знання, що панували на терені України до XX ст., та основні досягнення у галузі фізики та астрономії. Основну увагу приділено їх використанню при викладанні цих дисциплін.

Під час дослідження також вивчено історичні нариси з розвитку освітянських центрів [9–11], які містять відомості про наукові здобутки вчених у різних природничих галузях, зокрема фізики та астрономії.

Проведене автором опрацювання попередніх публікацій показало, що зв'язок релігії та уявлень про Всесвіт в Україні протягом XI–XVIII ст.ст. не було метою їх досліджень. Завдання нашої статті – детально проаналізувати вплив православного християнства на розвиток космологічних ідей в Україні протягом XI–XVIII ст.ст.

Джерельною базою для встановлення питання розвитку уявлень про будову Всесвіту до XX ст. слугували насамперед праці науковців-педагогів того часу [12–21].

Космологія – це наука про Всесвіт у цілому, його виникнення та еволюцію. На думку О.В. Огірко, релігія та наука – це два способи пізнання світу [4]. Завдяки ним формується інтегрований світогляд людини, вказується існування надприродного та матеріального світу. Формування та розвиток світогляду на теренах України брали початок з часів, коли людина ще не могла активно впливати на природу, була її невід'ємною частиною та повністю залежала від неї. Міфологічний характер стародавнього світорозуміння був пов'язаний із тим, що раціональне пояснення було можливим щодо окремих дій і явищ, але не могло забезпечити осмислення всього Всесвіту. Згодом найвищий авторитет здобули біблійні тексти, але у Святому Письмі не можна було знайти відповідей на конкретні питання щодо будови світу, властивостей речей, походження явищ. Для цього широко використовувалися праці античних авторів, передусім Арістотеля і Птолемея [22].

Візантійською культурою IX–X ст.ст. багато зроблено для систематизації знань, наявних в античній літературі [23, 24]. При цьому характерною рисою того часу було прагнення до цілісного знання, звідси і відсутність його чіткого розподілу за окремими науковими дисциплінами. У цей період відбувалася систематизація античних знань, яка задовольняла практичні потреби. Через Візантію у слов'янській світ також проникали твори православних богословів, численні переклади апокрифів, їх оригінальні інтерпретації та компіляції [12–18]. Серед українських мислителів були прихильники різних картин світу – геоцентричної та багатоярусної моделей, утворення Всесвіту відповідно до тексту Біблії тощо.

**Вплив християнської літератури на світосприйняття епохи Київської Русі.** У цей період відбувалося з одного боку – розповсюдження християнської літератури, з іншого – засвоєння знань, поширених у Візантії. На теренах України праці християнської спрямованості мали вагомий внесок на розвиток космологічних поглядів. Викладені в них концепції пояснювали появу та еволюцію Всесвіту відповідно до тексту Святого Письма. Мислителі Київської Русі уявляли собі будову світу по-різному, зокрема їм були відомі ідея плоскої Землі та геоцентрична модель Всесвіту Арістотеля–Птолемея [1, 2]. Планет було вісім. Давньоруською мовою вони називалися «заблудницями» (у перекладі з давньогрецької мови πλανήτης – той, що блукає. Саме ж слово «заблудниця» походить від дієслів «блукати» або «блудити»). Назви планет були транскрипцією імен грецьких богів: Єрмес (Гермес) – Меркурій, Арей (Арес) – Марс, Зеус (Зевс) – Юпітер, Кронос – Сатурн, Афродіта – Венера. До планет зараховували також Місяць і Сонце [25].

З XI ст., починаючи з М. Пселла, спостерігається більш критичне ставлення до християнських догматів. Михайло Пселл (1018–бл.1096) був візантійським політиком, істориком і філософом [26]. Його літературна спадщина охоплює право, медицину, історію, агрономію, риторику, філософію та математику. Він мав звання Іпата філософів і був першим керівником філософської школи в Константинополі. У своїй праці «Загальна наука» М. Пселл виклав уявлення про світобудову, згідно з яким Бог – творець природи, але остання підпорядковується власним законам. Він не задовольнявся поглядами Арістотеля і схилявся до неоплатонізму з його вченням про Єдине, Світовий Розум та Світову Душу. М. Пселл спирався на птолемеївську геоцентричну концепцію світобудови, уявляючи Всесвіт як сукупність обертальних небесних сфер навколо кулястої Землі [6, 25, 27].

Близьким до його поглядів був Симеон Сифа (друга половина XI – перша половина XII ст.) – візантійський мислитель і письменник. У своїй праці «Загальний огляд початків природознавства» він дотримувався теолого-неоплатонічного спрямування. Основними джерелами для написання цього трактату були праці Платона, Арістотеля та Птолемея. У передмові викладено ідеї щодо будови Всесвіту, форми небесних тіл і Землі, походження природних явищ тощо. Праця починається з опису сферичної форми Землі та її розміру. Потім викладено уявлення про сферу Місяця, Меркурія, Венери, Сонця, за якою по велінню Бога рухаються всі інші планети, Марс, Юпітер, Сатурн і нерухомі зорі. Останні світять від Сонця та власним світлом. У небесних сферах С. Сифа відмічав перевагу повітряного початку, а у зорях – вогневого. Далі йдуть відомості про небесний екватор, небесний меридіан, горизонт і нахил до нього осі світу, місячні та сонячні затемнення, кордони ойкумени (від м. Сіри (Китай) на сході до Іспанії на заході; від острова Фулі на півночі до екватора на півдні) [23].

В українських літописах, починаючи з XI–XII ст.ст., з'являються описи будови світу [3]. У той час їх існувало кілька варіантів. Серед них «Християнська топографія» Козьми Індикоплова [12], «Шестоднев» Іоанна (екзарх болгарський) [13], «Хроніка» Георгія Амартола [28]. Під впливом цієї літератури та внаслідок самостійного осмислення в Україні формувалися уявлення про будову світу. Астрономічну інтерпретацію космологічних ідей – систему Птолемея – було викладено в трактаті «Ізборник» 1073 р. [28]. Саме «Ізборники» 1073 та 1076 рр. були для мешканців Київської Русі одними з основних джерел знань. Вважається, що ці рукописи написано в Києві. За основу автори взяли твір візантійського походження. У першій частині «Ізборника» викладаються астрологічні поняття в їх критичному осмисленні. У другій частині – літочислення в різних народів, назви місяців. Дана праця містить основні положення «Метафізики» Арістотеля. За цим трактатом у центрі Всесвіту знаходиться Земля, навколо якої обертаються вісім небесних сфер, що пов'язані з рухомими тілами, які пересуваються на своїй орбіті – сфера Кроносу (Сатурну), Зеусу (Юпітеру), Арею (Марсу), Сонця, Афродіти (Венери), Єрмію (Меркурію), Місяцю та нерухомого неба. Остання також оберталась, але в протилежному руху планетних сфер. Деякі мислителі виділяли й дев'яту – зодіакальну сферу [28].

Козьма Індикоплов (VI ст.) був візантійським купцем і мандрівником. Припускається, що його трактат «Християнська топографія« [14] було перекладено на Русі в кінці XII – на початку XIII ст. У праці описано будову світу відповідно до Святого Писання: «повєда християнський переказ божественного писання та хотєніа являти всього світу образ» [14, с. 278]. Він дотримувався ідей богословів антіохійської школи і намагався викласти свої уявлення про систему світу, ґрунтуючись на біблійному тексті. В перший день Бог створив перше небо – вічне світло, в другий – видиме небо та відокремив води під твердю від вод, що знаходяться над твердю. Земля зв'язана з першим небом по широті. Два неба разом із Землею утворюють Всесвіт. К. Індикоплов вважав, що Земля має форму плоского прямокутника, оточеного океаном. Вона переходить у гору, за яку ховається Сонце. Рухами Сонця, Місяця та зір, небесними й атмосферними явищами керують окремо для того призначені ангели. На схилах Землі живуть різні народи. Небесне склепіння, тверде та прозоре, має форму намету. Намет – це перше небо. Друге небо – твердь, що має вигляд шкіри, натягнутої над першим небом. Сонце та Місяць розташовані нижче. Місяць світить своїм світлом, він не зникає, а ховає своє світло. Нерухомими є Земля та небо, а зорі, Сонце, Місяць – рухомі світила. Його світ – двоярусна споруда, нижній поверх якої займає сфера природи, а верхній відділений подвійним небом (твердь і вода та невидиме небо). К. Індикоплов порівнював Землю з Ноєвим ковчегом і старозавітною скинією. Він виступав проти авторитетів Арістотеля і Птолемея, які вважали, що Земля кругла, і наполегливо доводив, що Землю «немощно круглообразно порозумівати» [7, с. 8].

«Хроніка» [15] Георгія Амартола була завершена близько 867 р. Достеменно невідомо, де саме її було переведено на церковно-слов'янську мову, але згідно з однією із теорій це здійснено в Київській Русі XI–XII ст.ст. Трактат Г. Амартола складається зі вступу та чотирьох книг. У першій книзі викладено основні етапи історії від створення світу Богом до часів Александра Македонського.

Ці досить-таки примітивні уявлення про світобудову не були єдиними серед українського культурного загалу. Паралельно в Україні, як вже зазначалося, поширювалася праця Іоанна, екзарха болгарського, «Шестоднев», в якій астрономічна проблематика висвітлювалася з урахуванням досягнень античності [16, 29]. Вважається, що цей трактат було написано у Болгарії в кінці IX – на початку X ст.ст. Твір Іоанна являє собою поєднання «Шестодневу» Василія Великого, однойменної праці вченого-богослова Северіана Гевальського і трудів античних авторів, зокрема Арістотеля. Перші згадки про систему світу Іоанна містяться в літописі XIII ст. Згідно з трактатом, у перший день Бог створив небо і землю, у другий розділилася волога від тверді, в третій – води відокремилися від суходолу і виник рослинний світ, у четвертий – були створені небесні тіла та світила, в п'ятий – тварини, що живуть у воді, та птахи, а у шостий – наземні тварини і людина. Цей зміст концепції виникнення світу відповідає Святому Писанню. Іоанн скомпілював працю, яка значно ближче підійшла до наукового пояснення світобудови, ніж трактат К. Індикоплова. Якщо останній твердив про плоску форму Землі, то болгарський мислитель наводить думку Арістотеля, який обґрунтував її кулястість. Відповідно до «Шестоднева», кулеподібна Земля розташована у центрі сферичного небесного склепіння, з яким пов'язані кілька концентрично розташованих рухомих кіл. До кіл прикріплено Сонце та Місяць, п'ять планет («плаваючих зір»), які здійснюють петлеподібні рухи, та нерухомі зорі. Сонце, Місяць, зорі та все суще має форму кулі. Сонце рухається по підземній підлозі та вищеземній. За рік воно проходить 12 зодіакальних сузір'їв, рухаючись «зодіастим», або «живоносним» колом. У цьому «Шестодневі» містилося багато астрономічних відомостей: дані про розмір небесних тіл, пояснення рівнодення та сонцестояння, зміни пір року, відхилення тіні в Південній півкулі й уявлення про кліматичні пояси Землі. Тут викладалися погляди на будову світу Птолемея та К. Індикоплова. Цю книгу можна вважати провідником птолемеївських ідей на Русі.

Поширення космографічних ідей «Християнської топографії» К. Індикоплова, «Хроніки» Г. Амартола та «Шестоднева» Іоанна дає підстави вважати, що астрономічні уявлення в середньовічній Україні розвивалися в двох напрямах: у формі тлумачення відповідних текстів Святого Письма і по лінії засвоєння астрономічних знань Стародавньої Греції. Обидва ці напрями ґрунтувалися на християнських засадах, другий, спираючись на античні традиції, допускав ширший погляд на світ, і в поясненні його будови та явищ виходив із закономірностей природи. Ці праці були дуже популярними серед мислителів Київської Русі, вони бережно зберігалися та переписувалися протягом багатьох століть. Повних списків «Шестоднева» налічується понад 50, а «Християнської топографії» збереглося понад ніж у 90 списках.

**Вплив християнських догматів на космологічні моделі на теренах України в період XV – XVIII ст.ст.** Переламним етапом в історії космології є XV–XVI ст.ст., коли створено геліоцентричну систему світу. В Україні в той час були поширені два астрономічні трактати – «Космографія» [30] Іоанна Сакробоско та «Шестокрил» Імануель-бар-Якоба [8].

Іоанн Сакробоско, або Джон із Галіфакса (бл. 1200–1256) був англійським астрономом і математиком. Після закінчення Оксфордського університету він викладав у Паризькому університеті. Один із перших став широко використовувати переведену астрономічну літературу арабських мислителів. І. Сакробоско написав трактат «Про сферу», в українському перекладі відомий під назвою «Космографія» [30]. Відповідно до його поглядів, Земля, поверхня води та весь Всесвіт мали кулеподібну форму. Кривизну земної поверхні він пояснював різницею у часі затемнень на Сході та Заході, а також відмінністю у видимості зір. Кривизну поверхні води І. Сакробоско доводив через той факт, що людина, яка стоїть на основі корабельної щогли, не бачить предмети, видимі людині, яка стоїть зверху цієї щогли. Крім того, оскільки вода – однорідна речовина, кожна частинка якої повинна мати властивості цілого, і якщо водяна крапля кругла, то й уся воднева маса кулеподібної форми. Рух небесних світил пояснювався самообертанням сфер, а не надприродними силами. Систему Всесвіту «Космографія» подає за Птолемеєм [30]. Небесних кіл дев'ять. Концентричне розташування сфер Птолемея дається в ній наочно: «Усіє небеса один ув одном, як цибуля». Тим часом «Земля бо у самой середині неба, а не виходить нікодиже із місця свого» [30, с. 178]. Це обумовлено, по-перше, її вагою: оскільки вона важкий елемент, а усім важким тілам властиво прямувати до центру Всесвіту. По-друге, згідно з трактуванням І. Сакробоско, якщо Земля не у центрі, з її областей, розташованих ближче до небесного склепіння, було б неможливо бачити його середину, тобто небесний екватор, і шість знаків зодіаку, які однаково видно з будь-якої точки земної поверхні. Відповідно до «Космографії», на дві третини землю покриває вода. Над ними гуляє вітер, породжений взаємодією землі, вологи і тепла. Вогонь названий колесом вітру [30].

«Космографія» І. Сакробоско була доволі популярна і збереглася у багатьох манускриптах. Вона використовувалася як базовий твір при вивченні астрономії у середньовічних університетах. Як відмічає І.В. Паславський [31], відстоювання автором «Космографії» кулястості Землі було значним кроком уперед порівняно з християнськими поглядами, за якими форма Землі визнавалась плоскою.

Іншим розповсюдженим у той час на теренах України астрономічним трактатом був «Шестокрил» єврейського вченого XIV ст. Імануель-бар-Якоба. Його переклад і поширення пов'язують з іменем київського мислителя (родом з м. Кафа) Захарії (Схарія), який у середині XV ст. був особистим астрологом київського князя Михайла Олельковича. Він вивчав птолемеївську систему світу, перекладаючи та розповсюджуючи «Шестокрил» [31], який можна вважати практичним посібником з астрономії. Він являє собою шість місячних таблиць для використання в астрономічних обчисленнях: «заведі пальцами от ширіни страніци і от должини страніци, штоб ся на одной строце споткалі» [31, с. 183].

Переклад і появу на теренах України в другій половині XV ст. праць «Космографія» та «Шестокрил» можна вважати новим етапом у розвитку астрономічних уявлень нашого народу. Ці трактати несли в освічені верстви вчення про Всесвіт Арістотеля–Птолемея, руйнуючи тим самим християнські уявлення про світобудову. На противагу попереднім поглядам, сонячні та місячні затемнення вже не трактувалися як божественне покарання. Їх появу навчилися вираховувати, передбачати та пояснювати законами природи.

Нагромадженню фізико-математичних знань в Україні сприяла діяльність українських гуманістів кінця XV – початку XVI ст. Після здобуття вищої освіти в західноєвропейських університетах Відня, Падуї, Болоньї, Венеції, Риму, Кракова українці переносили на національний ґрунт ідеї епохи Відродження. Творчість нової генерації гуманістів припала на кінець XVI – першу половину XVII ст.ст. Її представники в своїй просвітницькій діяльності гуртувалися навколо культурно-освітніх осередків, найзначнішим серед яких була Острозька академія, яка поєднала давні українські та греко-візантійські освітні традиції з досягненнями європейської освіти. В її книжному зібрані були твори Арістотеля, Вергілія, Ксенофонта, Сенеки, Цицерона, Ф. Петрарки, П. Мануція, І. Сакробоско, Т. Браге та ін. [32]. Тут читались «сім вільних наук», зокрема математика, астрономія, філософія. Учні колегії користувалися працями з математики, астрономії, філософії, фізики, написані переважно латиною. Серед них «Космографія» І. Блеу, яка містила відомості про систему світу Коперника і була відома в російському перекладі Є. Славинецького 1645–1647 рр. під назвою «Зерцало усього Всесвіту…», також «Астрономічний календар за 1506 р.", "Фізика та сферика" (1593 р.) та інші [33].

На кінець першої чверті XVII ст. Київ утвердився як провідний православний церковний і культурно-освітній центр. Його авторитет зміцнився завдяки утворенню Братської школи, активній друкарській роботі Лаврського освітницького гуртка та відновленню православної митрополичої кафедри. У цей час у місті почав діяти Києво-Могилянській колегіум, із вихованців якого було створено потужний прошарок освіченого духовенства, яке дбало про поширення освіти. За структурою, обсягом і змістом курсів колегіум відповідав вимогам, що ставилися перед тогочасною західноєвропейською вищою школою, але викладачами були православні священники.

У 1627 р. вийшов «Катехізис» Л. Зизанія [34], в якому викладено християнські уявлення щодо появи Всесвіту. Слід зауважити, що це XVII ст., коли в Західній Європі в університетах викладалися модель Коперніка, ідеї Джордано Бруно та Декарта. Лаврентій Зизаній (60-ті рр. XVI ст. – 1628 р.) був українським церковним діячем, мовознавцем, письменником, перекладачем і богословом. Він викладав у Львівській, Берестейській і Віленській братських школах, перекладав із грецької на церковнослов'янську мову в Києво-Печерській лаврі. В «Катехізисі» на староукраїнській мові він писав «о крузях небесних, і о планітах, і о зодіях, і о затмені сонця, про грім і молнію, про перуна та комети, і про інші зорі» [34, с. 92]. В своєму трактаті Л. Зизаній намагався дати пояснення багатьох явищ природними причинами. Московський патріарх Філарет не дозволив випустити твір і постановив спалити його! Випадково вціліло декілька примірників. На звинувачення з боку московських опонентів Л. Зизаній говорив, що виклав «ведомости ради, чтобы человек ведал, яко то есть», а не для того, щоб показати, що «звездами правится житию нашему» [34, с. 93]. Брат Л. Зизанія – Стефан Зизаній (бл. 1570–1621) висловлював ідеї множинності та безмежності світів (аналогічно до поглядів Джордано Бруно): «Абовім вся земля, на которой живем, як єдина точка в посередку неба, а пред ся колкоє мают множество; а небеса небеснії єще болше безмірную мают лічбу» [18]. Нагадаємо, що італійський філософ Джордано Бруно за пропагування ідей про нескінченність Всесвіту та про безліч світів був визнаний єретиком і був спалений у Римі. Твердження С. Зизанії в кінці XVI ст. були надто сміливими і межували з релігійним вільнодумством.

Наприкінці XVII ст. Києво-Могилянський колегіум набув статусу академії. Тут було запроваджено нормативне викладання богослов'я. Навчання у вищих класах академії тривало шість років і передбачало передусім засвоєння філософії за два роки та богослов'я – за чотири. Філософія поділялася на «натуральну філософію» (з подальшим удосконаленим вивченням математики) і метафізику. Перша включала основи природничих знань – «фізику», до складу якої входили наука про Всесвіт, метеорологія та фізіологічна психологія в зв'язку із зоологією. Метафізика розглядала надприродні чинники світових явищ, їх першопричини і загальні принципи відповідно до віри в створення світу Богом [9, 10].

Курси філософії, які читалися тут у XVII–XVIII ст.ст., у цілому були подібними до тих, що викладалися в провідних європейських університетах. Однак авторитетними залишалися християнські православні ідеї щодо виникнення Всесвіту. Сучасні дослідники філософської спадщини Києво-Могилянської академії розрізняють у філософській думці академії два напрямки. Один із них науково-освітній. Представники його головним вважали розвиток науки, освіти, ремесел, мистецтв і виховання. Це була так звана арістотелівсько-природознавча раціоналістична лінія. До неї належали І. Гізель, Й. Кроковський, Ф. Прокопович, Г. Кониський, М. Козачинський та ін. Професори академії С. Клешанський, С. Яворський, І. Поповський, не задовольняючись теорією Арістотеля–Птолемея, спрямували свої пошуки на інші теорії будови світу. Вони ще не визнавали вчення Коперника єдино правильним, проте під час його викладання застосовували таблиці, креслення, що давало змогу слухачам академії самим розібратися в достовірності тієї чи іншої теорії.

Інокентій Гізель (бл. 1600–1683) був українським релігійним діячем, богословом, філософом й істориком. Він навчався в Колегіумі Петра Могили, вивчав філософію, богослов'я, право й інші науки в Замойській академії, в університетах Німеччини й Англії [19]. І. Гізель був архімандритом Києво-Печерської лаври та ректором Києво-Братської колегії. У своїх текстах він використовував матеріали античності, патристики та схоластики; орієнтувався на ідеї Г. Галілея, М. Коперника та Д. Кардано, був прихильником арістотелізму. Його філософський курс «Твір про всю філософію», прочитаний в академії в 1645–1647 рр., містить знання з усіх розділів тогочасної філософії. Хронологічно він є першим курсом натурфілософії, прочитаним в академії. Основне місце в ньому відведено натурфілософії, де проводиться ідея єдності й однорідності матерії землі й неба, незнищуваності матерії, її переходу від однієї форми до іншої. І. Гізель поряд з геоцентричною системою світу, прихильником якої він був, аналізував систему Коперника, в такий спосіб в Україні вперше згадується ім'я М. Коперника. І. Гізель пояснював учням причини затемнення Місяця, підкреслював, що зорі світяться власним світлом, а планети – відбитим.

Йоасаф Кроковський (1650–1718) – український церковний діяч і ректор Києво-Могилянської колегії. Тут він читав курс філософії «Диспут з логіки, є погоджена побудова арістотельського Органона, викладена в колегіумі Києво-Могилянській» [35]. Курс ділився, як у Арістотеля, на три частини: логіку, фізику (натурфілософію) та метафізику. Й. Кроковський розглядав натурфілософію як головну частину при навчанні. У курсі він виклав своє розуміння Всесвіту, його будову, пояснив земні та небесні явища тощо. Мислитель писав: «Ми повинні йти за Арістотелем, але не сліпо» [35].

Мануїл Козачинський (1699–1755) був ученим-гуманістом, викладачем, письменником. Після закінчення Києво-Могилянської академії працював у Сербії, а з 1739 р. – повернувся викладати до alma mater. У 1743 р. прочитав курс «Філософія Арістотелева…», який було опубліковано в 1745 р. [20].

Феофан Прокопович (1677–1736) навчався в Києво-Могилянській і Римській академіях, слухав лекції в німецьких університетах Галле, Лейпцига, Йєни, Кенігсберга. У 1705–1716 рр. викладав усі вищі науки в академії. Перший почав ґрунтовно знайомити студентів із вченням Декарта, Локка, Бекона, дав пояснення системи Коперника. Про фізику він писав: «вона, запліднюючи всі мистецтва, подає велетенську користь родові людському» [21, с. 115].

У своїй праці «Натурфілософія, або Фізика» [21, с. 113–502] він дав визначення світу. За ним світ – структура, що складається з неба, землі та інших елементів, що знаходяться між ними, або світ – це «порядок та розташування всього, що зберігається богом та завдяки богу» [21, с. 283]. Ф. Прокопович знайомив з усіма поширеними на той час уявленнями про Всесвіт. Спершу викладалась система світу Птолемея, де вчений вказував на те, що цю світобудову запропонував Піфагор. Проте, оскільки Птолемей, «будучи видатним математиком, пояснив свою систему більш детально, обґрунтувавши її різними аргументами» [21, с. 286], вона носить назву – система світу Птолемея.

Наступною системою, що викладена у Ф. Прокоповича, була модель Коперника. Він зазначав, що дана теорія є недостатньою для пояснення багатьох складних питань астрономії. Крім логічних аргументів, від цієї системи особливо відштовхували людський розум свідчення Святого Письма, згідно з яким не Земля рухається, а Сонце. Далі він знайомив із системою світу Тихо Браге.

Не дивлячись на те, що в своєму курсі Ф. Прокопович викладав різноманітні системи світу, основною ідеєю залишалось його божественне творіння. «Найбільшим і найочевиднішим доказом того, що світ виник не випадково, не внаслідок сліпого змішування атомів, а був створений якимось наймудрішим і одночасно наймогутнішим творцем, – писав він, – є сама величина цього світу, його краса, різноманітність, будова, різні роди речей, що містяться в ньому і властиві окремим формам» [21, с. 291]. Згідно зі Святим Письмом, світ не існував вічно, «спочатку було створено небо й землю» [21, с. 296]. І на це саме джерело Ф. Прокопович посилається при твердженні, що світ є один, оскільки в Біблії постійно згадується і описується лише один світ.

Григорій Кониський (1717–1795) був українським і білоруським релігійним діячем, письменником і викладачем [36, 37]. Після закінчення Київської Академії він залишився в ній працювати, з 1755 р. – жив у Могильові. В своїх лекціях Г. Кониський знайомив учнів з різними моделями Всесвіту (від ідей античного світу до вчення мислителів Нового часу) та останніми досягненнями з астрономії. Наслідуючи традиції Ф. Прокоповича, він поширював ідеї геліоцентризму в Україні, детально виклав вчення Коперника з переліком імен багатьох учених попередників і послідовників цієї моделі, знайомив із вихровою теорією Декарта. При викладанні останньої Г. Кониський припустив можливість відкриття нових планет, вічність матерії, нескінченність світу та множинність світів [38].

**Висновки.** Підсумовуючи вищенаведене, можна зробити висновок, що на теренах України в досліджуваний період християнська література мала авторитетний вплив на розвиток уявлень про виникнення та еволюцію Всесвіту. При цьому широко розповсюджувались, перекладались і зберігались численні трактати закордонних філософів і богословів. Переведенні тексти інтенсивно тиражувались і записувались у добірки натурфілософського змісту. На теренах України були відомі всі моделі Всесвіту, що популяризувалися у Західній Європі. Викладачі існуючих на той час колегіумів (представники духовенства) знайомили студентів з моделями Всесвіту Птолемея, Коперника, Декарта, Канта–Лапласа. Однак божественне створення світу було невід'ємною складовою уявлень про Всесвіт.